\documentclass[reprint, a4paper, twocolumn,
superscriptaddress,
groupedaddress,
 amsmath,amssymb,
 cond-mat,
pra,
floatfix,
]{revtex4-2}

\usepackage{graphicx}
\usepackage{dcolumn}
\usepackage{bm}

\usepackage{hyperref}
\begin{document}

\preprint{APS/123-QED}

\title{Magnetic Fluctuations in Niobium Pentoxide}
\author{Y. Krasnikova}
\affiliation{Superconducting Quantum Materials and Systems Center (SQMS),
Fermi National Accelerator Laboratory, Batavia, IL 60510, USA}

\author{A. A. Murthy}
\affiliation{Superconducting Quantum Materials and Systems Center (SQMS),
Fermi National Accelerator Laboratory, Batavia, IL 60510, USA}

\author{F. Crisa}
\affiliation{Superconducting Quantum Materials and Systems Center (SQMS),
Fermi National Accelerator Laboratory, Batavia, IL 60510, USA}

\author{M. Bal}
\affiliation{Superconducting Quantum Materials and Systems Center (SQMS),
Fermi National Accelerator Laboratory, Batavia, IL 60510, USA}

\author{Z. Sung}
\affiliation{Superconducting Quantum Materials and Systems Center (SQMS),
Fermi National Accelerator Laboratory, Batavia, IL 60510, USA}

\author{J. Lee}
\affiliation{Superconducting Quantum Materials and Systems Center (SQMS),
Fermi National Accelerator Laboratory, Batavia, IL 60510, USA}

\author{A. Cano}
\affiliation{Superconducting Quantum Materials and Systems Center (SQMS),
Fermi National Accelerator Laboratory, Batavia, IL 60510, USA}

\author{D. M. T. van Zanten}
\affiliation{Superconducting Quantum Materials and Systems Center (SQMS),
Fermi National Accelerator Laboratory, Batavia, IL 60510, USA}

\author{A. Romanenko}
\affiliation{Superconducting Quantum Materials and Systems Center (SQMS),
Fermi National Accelerator Laboratory, Batavia, IL 60510, USA}

\author{A. Grassellino}
\affiliation{Superconducting Quantum Materials and Systems Center (SQMS),
Fermi National Accelerator Laboratory, Batavia, IL 60510, USA}

\author{A. Suter}
\affiliation{Laboratory for Muon Spin Spectroscopy, Paul Scherrer Institute, CH-5232 Villigen PSI, Switzerland}

\author{T. Prokscha}
\affiliation{Laboratory for Muon Spin Spectroscopy, Paul Scherrer Institute, CH-5232 Villigen PSI, Switzerland}

\author{Z. Salman}
\affiliation{Laboratory for Muon Spin Spectroscopy, Paul Scherrer Institute, CH-5232 Villigen PSI, Switzerland}

\date{\today}

\begin{abstract}
Using a spin-polarized muon beam we were able to capture magnetic dynamics in an amorphous niobium pentoxide thin film. Muons are used to probe internal magnetic fields produced by defects. Magnetic fluctuations could be described by the dynamical Kubo-Toyabe model considering a time-dependent local magnetic field. We state that observed fluctuations result from the correlated motion of electron spins. We expect that oxygen vacancies play a significant role in these films and lead to a complex magnetic field distribution which is non-stationary. The characteristic average rate of magnetic field change is on the order of 100~MHz. The observed dynamics may provide insight into potential noise sources in Nb-based superconducting devices, while also highlighting the limitations imposed by amorphous oxides. 

\end{abstract}

\maketitle


\begin{figure}[!htb]
\includegraphics[width=0.5\textwidth]{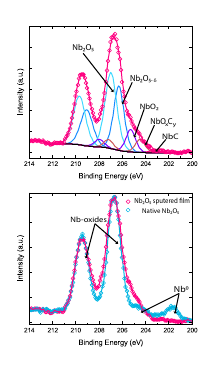}
\caption{\label{fig:XPSoxides} Upper panel: XPS data analysis for Nb$_2$O$_5$ DC sputtered film. Lower panel: XPS data comparison for native surface Nb oxides and Nb$_2$O$_5$ DC sputtered film, both are from the same batch as used for measurements at beam-facility at PSI.}
\end{figure}

\begin{figure*}[!htb]
\includegraphics[width=1\textwidth]{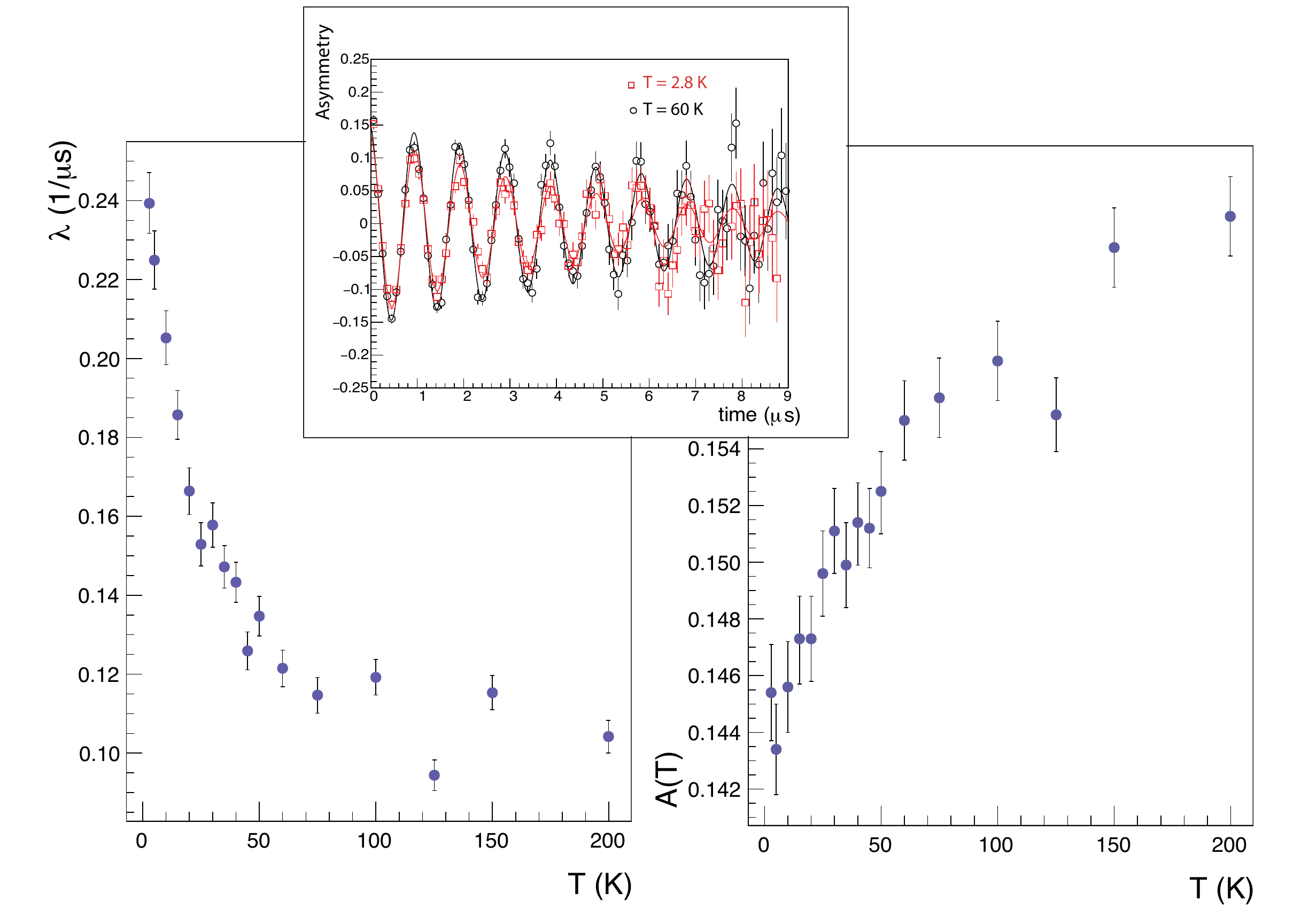}
\caption{\label{fig:TF}  $\mu$SR data in transverse magnetic field 75~Gauss (TF) in Nb$_2$O$_5$, $E=9$~keV. Left panel: temperature dependence of depolarization rate. Right panel: temperature dependence of asymmetry coefficient. Insert: examples of raw data.}
\end{figure*}

\begin{figure}[!htb]
\includegraphics[width=0.48\textwidth]{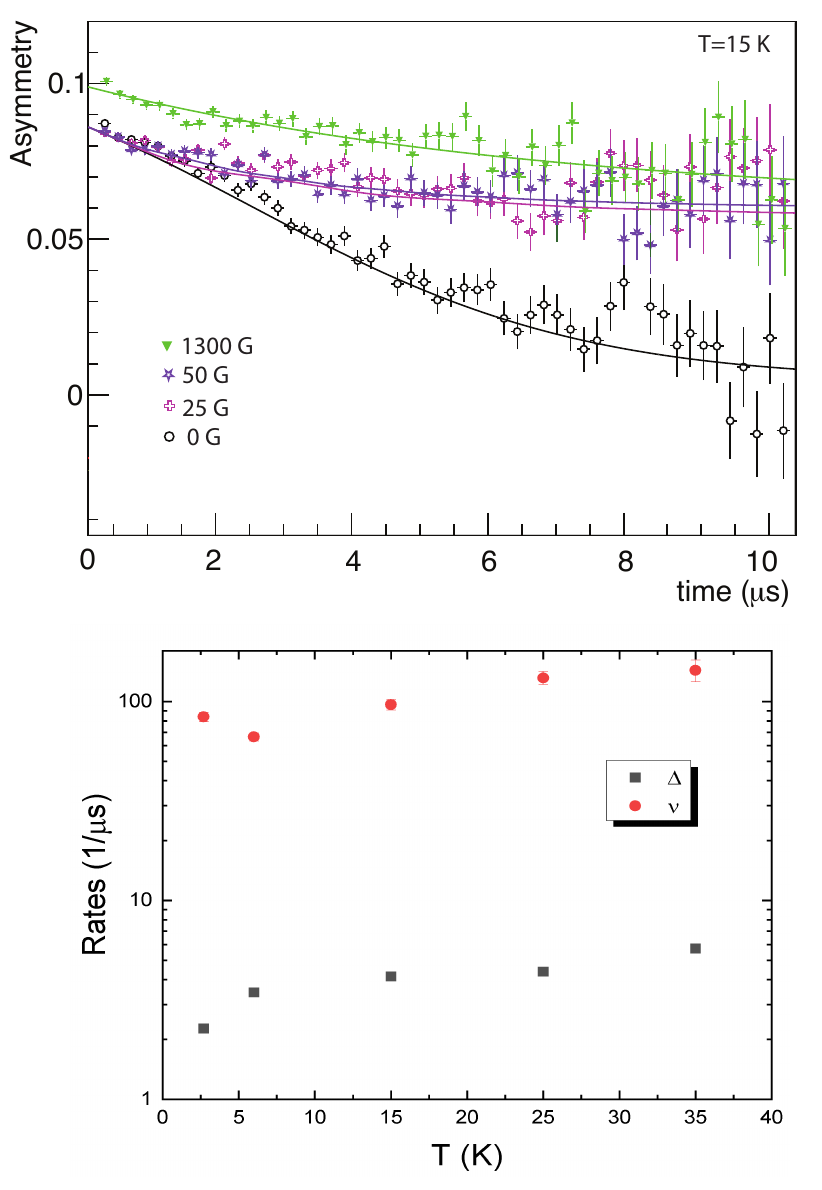}
\caption{\label{fig:LF} $\mu$SR data in longitudinal magnetic field (LF) in Nb$_2$O$_5$, $E=9$~keV.  Upper panel: Asymmetry function in different magnetic fields at fixed temperature. Lower panel: results of global fitting for asymmetry function at different temperatures, taking dynamic Kubo-Toyabe function, $\Delta$ -- static and $\nu$ -- the dynamic rate for local magnetic field.}
\end{figure}

\begin{figure}[!htb]
\includegraphics[width=0.5\textwidth]{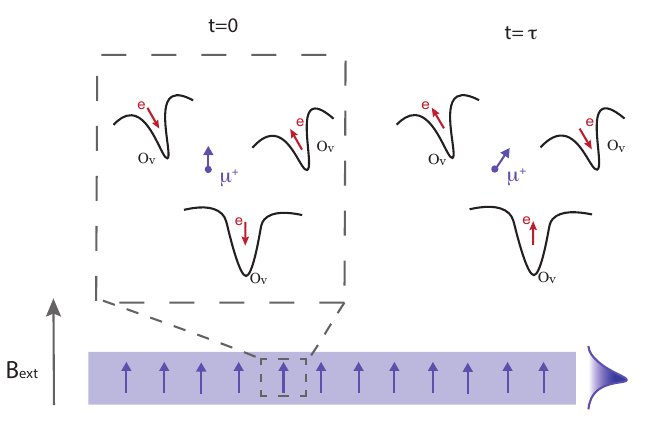}
\caption{\label{fig:wells_slides} Naive model to describe polarized muon ($\mu^{+}$) sitting in complex of vacancies (O$_{v}$), nuclear moments are not shown for simplicity. The main assumption that vacancy works as defect and forms trapping potential well, electrons (e) which are sitting inside potential well might have its own spin polarization. This polarization could have time dependent component based on our $\mu$SR experimental data, which means that electrons' spins are not frozen.}
\end{figure}

The performance of superconducting qubits is currently limited by a wide variety of loss mechanisms arising from defects and impurities present at material interfaces and surfaces in these devices \cite{Krantz2019}. It is often a challenge to distinguish between various factors when the experimental probe or experimental method itself is responsive to many degrees of freedom. In the case of niobium-based qubit devices, the niobium surface oxide (Nb$_2$O$_5$) that spontaneously grows in ambient conditions serves the major source of microwave loss \cite{Romanenko_Schuster2017, Romanenko2020, Bal2023}. This loss has been previously linked to the presence of oxygen vacancies \cite{Greener1961, Proslier2011}, as each oxygen vacancy likely introduces an unpaired electron for Nb$_2$O$_5$ \cite{Streiff1971}. Further, antiferromagnetic behavior has been observed in oxygen-deficient niobium oxides \cite{Cava1991}. Another one reason the magnetic footprint in niobium pentoxide is worth exploring is the surface spin model, which was actively discussed for Nb-based SQUID magnetometers \cite{Sendelbach2008} and lately for frequency tunable resonators \cite{Graaf2020}.
Oxygen deficiency by itself might cause magnetism, even ferromagnetism in dielectric oxide, in cases where it may not be theoretically expected \cite{Venkatesan2004}.

In this Letter, we describe experimental results of low-energy muon spectroscopy as a probe of local magnetism in amorphous niobium pentoxide with various applied magnetic fields and different temperature settings.
$\mu$SR (muon spin resonance/rotation/relaxation) is a technique that has similarities to both electron spin resonance (ESR) and nuclear magnetic resonance (NMR). It allows for measuring dynamical phenomena occurring on time scales ranging from 10$^{3}$~s$^{-1}$ to 10$^{12}$~s$^{-1}$ \cite{Blundell2021}. The muon lifetime is about 2.2~$\mu$s and decaying positrons are emitted along the muon spin direction. As the spin polarization of each muon is impacted by the local magnetic fields present within a solid, the positron counts as a function of time provides insight into the spin depolarization of muons. In these experiments the asymmetry function, which is directly proportional to the polarization function, is commonly used to describe response to the local magnetic field.
The Paul Scherrer Institute has a unique spectrometer operated with a decelerated muon beamline down to keV range. This type of low-energy $\mu$SR (LE-$\mu$SR) is advantageous as it offers the ability to examine thin film samples. However, in comparison with bulk $\mu$SR, although beam is quasi-continuous, beam deceleration leads to an average of about 1800 events per second. Thus, repeated measurements over several hours are required to collect a sufficiently large signal to noise ratio, especially for longitudinal field. Nonetheless, this technique provides unique value in that allows for observing phenomena on time scales that are not accessible with comparable methods such beta-NMR \cite{Kiefl2000}, which provides dynamical information from longer time scales. This capability to detect dynamical phenomena across wide range of frequencies combined with the sensitivity to weak magnetic fluctuations makes this technique highly valuable for probing the dynamics at play during superconducting qubit operation. 

We have performed LE-$\mu$SR studies on amorphous Nb$_2$O$_5$ to better understand why this oxide limits the performance of superconducting qubit devices. A thin film of Nb$_2$O$_5$ was prepared using DC magnetron sputtering. X-ray photoelectron spectroscopy (XPS) was used to examine the chemical composition of these films as seen in Fig.~\ref{fig:XPSoxides}. From the Nb 3d spectra, we observe that although a majority of the Nb atoms in the oxide layer have a charge state of 5+, which corresponds to Nb$_2$O$_5$, we also observe signal that corresponds to Nb with lower charge states. We attribute this to the presence of oxygen vacancies in the oxide and label the spectral signal as Nb$_2$O$_{5-x}$. Nonetheless, through comparison with similar measurements performed at the surface of a Nb film, we observe that the grown Nb$_2$O$_5$ film is chemically very similar to the native Nb$_2$O$_5$ that forms at the surface of Nb. Namely, we observe alignment in the positions of Nb 3d peaks and a similar broadening in the peaks due to the presence of oxygen vacancies. The Nb0 charge state peak results from photoelectrons that are generated in the Nb film that lies beneath surface oxide in the sample. 
Further, a depth profile using time of flight secondary ion mass spectrometry (ToF-SIMS) measurements suggests that the stoichiometry of the Nb$_2$O$_5$ thin film sample remains fairly constant. This suggests that the observed oxygen vacancies are dispersed throughout the oxide. We do observe the presence of carbon in the top 5nm of Nb$_2$O$_5$ thin film, but this signal decays within the film and is attenuated within the depths that we probe with LE muons.

The low-energy muon facility provides the opportunity to vary the beam energy between 1 to 30~keV. In the case of this sample, this roughly corresponds to a range between 10 to 300~nm in thickness, depending on the energy and density of the material. No significant difference was observed for this sample between 3, 9 and 18 keV (20, 55 and 95 nm) \cite{Suppl}, which further supports the idea that this film is highly homogeneous. Based on this insight, the mean depth was selected for implanting muons and all further measurements were performed using an average beam penetration depth of $\sim$55 nm (9~keV). By choosing this implantation energy we can be confident that the signal corresponds to dynamical phenomena present in the Nb$_2$O$_5$ film as opposed to effects present at the vacuum interface or the substrate.

Two magnetic field configurations were used relative to muon spin. Assuming that the magnetic field at muon site $B_{loc}$ is homogeneous and parallel to the $Z$ axis The polarization functions can be described as:
\begin{equation}\label{eq:1}
\begin{split}
    P_{Z}(t)=P_{Z} \exp{(-\lambda_{Z}t)},\\
    P_{X}(t)=P_{X} \exp{(-\lambda_{X}t)}\cos({\gamma_{\mu}B_{loc}t}),
   \end{split} 
\end{equation}
$\lambda_{X}$ and $\lambda_{Z}$ are longitudinal and transverse relaxation rates.

The $\mu$SR measurements were taken in zero field (ZF), transverse field (TF) and longitudinal field (LF). The base temperature is 2.8~K.
We used musrfit to analyze the $\mu$SR data \cite{Suter2012musrfit}.

In the case of the transverse field polarization function:
\begin{equation}
\begin{split}
P_{x}(t)=A(T)\exp{(-\lambda t)}\cos{(\gamma_{\mu} B t + \phi)},
\end{split}
\end{equation}
$\lambda$ -- depolarization rate, $A(T)$ -- asymmetry parameter dependent on temperature, $\gamma_{\mu}$ -- muon gyromagnetic ratio, $B$ -- magnetic field, $\phi$ -- frequency and phase correspondingly defining muon precession in the magnetic field. The results of fitting are shown in Fig.~\ref{fig:TF}.

In transverse magnetic field, muons start to precess. This precession frequency is dependent on the external magnetic field. If the sample is magnetic, a change in the asymmetry will be observed. The asymmetry is plotted as a function of temperature. Based on this plot, we do not observe an obvious footprint associated with a second order transition. However, we observe that magnetic fluctuations appears to be significant below 40~K (Fig.~\ref{fig:TF}). Similarly, the depolarization rate is also temperature-dependent. But the most significant change is observed in the asymmetry, which suggests that there is a finite density of magnetic centers in the film about non-zero amount of magnetic centers. These centers are capable of depolarizing muons and introducing local magnetic fields, which can be referred to as the internal magnetic field. In the insert for Fig.~\ref{fig:TF} examples of raw data for captured for two temperatures $T=2.8~$K and $T=60~$K is presented. In this case, the amplitude differences are visually distinguishable. 

The most common model to describe magnetic fluctuations presence is the dynamical Kubo-Toyabe model \cite{Yaouanc2011}. It is a mean-field approximation that takes into account both static and dynamic input in the depolarization function. The rate of fluctuations in this model represents an average rate of collisions (fluctuations). Each muon sitting in the sample will face depolarization due to a stochastically varying field that is described by multiple "collisions". The hopping rate $\nu$ in the model represents time between collisions. For fast fluctuations ($\nu/\Delta >> 1$), there exists an analytical solution \cite{Blundell2021}. In the case of medium regime, an asymmetry function can only be numerically solved. In our case, muons are sensitive at rates corresponding to rate 100~MHz, although this can be extended by an order of magnitude to more closely match the operating frequencies of superconducting qubits. Noise sources measured in this regime can potentially serve as sources of decoherence in qubits.

For components of local field:
\begin{equation}
    \overline{B_{\text{loc}}(t_{0})B_{\text{loc}}(t_{0}+t)}=\overline{(B_{\text{loc}})^2}\exp{(-\nu\lvert t \rvert)},
\end{equation}
$\nu$ -- correlation frequency of the random local field $B_{\text{loc}}$.

Polarization could be defined in case of dynamic input presence by integral equation:
\begin{equation}
\begin{split}
P_{z}(t)=P_{z}^{\text{stat}}(t)\exp({-\nu t})+\\
+\nu \int_{0}^{t}P_{z}(t-t'){P_{z}^{\text{stat}}(t')\exp({-\nu t'})dt'},
\end{split}
\end{equation}
In case when $\nu=0$ polarization is defined by static Kubo-Toyabe function :
\begin{equation}
P_{z}^{\text{stat}}(t)=A\left( \frac{1}{3}+\frac{2}{3}\exp{(-1/2\Delta^2t^2)}(1-\Delta^2t^2) \right),
\end{equation}
$\Delta$ -- static field distribution parameter.

To describe our data, we used a two subsystem approach. We assumed that at high temperatures the asymmetry behavior is largely defined by effects arising from nuclear subsystem, while at lower temperatures we assumed that the asymmetry depends on the sum of contributions from both the nuclear and electronic system \cite{Suppl}. The numerical fitting results are shown in Fig.~\ref{fig:LF}. 

An important detail of muon spectroscopy is that free unpaired electrons will not depolarize muons. The detected non-zero hopping rate ($\simeq$100 MHz) in dynamical Kubo-Toyabe model is associated with the presence of correlated objects. It is most likely electrons are interacting, potentially forming clusters with non-zero spin and charge density. Further theoretical work is necessary to describe flux and charge noise connection for this system. A simplified model of the environment that the muon 'sees' local magnetic field in amorphous niobium pentoxide  is provided in Fig.~\ref{fig:wells_slides} and more complex models are necessary to capture the full physics of these interactions. For instance, LF measurements in 1300~G magnetic field still lead to a decrease in the asymmetry, however in this field, nuclear moments are decoupled from muon spin. The drop in asymmetry in this case can be best explained by incorporating the contribution of electrons.

The main conclusion from this study is that electron spins in amorphous niobium pentoxide are not static even in the case where they may be trapped. However, freezing of magnetic moments were not detected at any temperature. Additionally, we observe dynamic depolarization and believe it arises due to electrons. These electrons are not free electrons. Rather, they are correlated, without long-range order formation. The characteristic temperature when these interactions become significant is $T\approx 35$~K (3~meV). 

The presence of both dynamic and static magnetic fields in the native surface oxide of niobium could be crucial for superconducting devices due to proximity effect. Static moments may lead to depairing of Cooper pairs and magnetic moments dynamics may contribute to the flux noise. 

\begin{acknowledgments}
This material is based upon work supported by the U.S. Department of Energy, Office of Science, National Quantum Information Science Research Centers, Superconducting Quantum Materials and Systems Center (SQMS) under contract no. DE-AC02-07CH11359. This work made use of the SmuS Facility of Paul Scherrer Institute. The authors thank for valuable discussions members of SQMS Center collaboration, in particular Ziwen Huang and Graham Pritchard. Valuable conversation authors had with Lara Faoro and Lev Ioffe and authors are thankful for their independent expert opinion.
\end{acknowledgments}

\nocite{*}

\bibliography{main}

\end{document}


\preprint{APS/123-QED}

\title{Supplemental materials}

\begin{abstract}
\end{abstract}

\maketitle


{\textbf{Film Deposition conditions}} 
\newline
Nb$_2$O$_5$ films were prepared using reactive sputter deposition. A Si wafer is loaded inside an AJA ATC 2200 sputtering system with a base pressure better than $10^{-8}$ Torr. DC magnetron sputtering was performed using a 3-inch diameter Nb target with a metals basis purity of 99.95\% with an Ar flow rate of 30 sccm, an O$_2$ flow rate of 20 sccm, and partial pressure of 3.5 mTorr at room temperature. A sputtering power of 600 W was used and the substrate was rotated at 20 rpm. These conditions resulted in a Nb$_2$O$_5$ deposition rate of approximately 5 nm/min.

{\textbf{Film XPS Characterization}}
\newline
X-ray photoelectron spectra were collected using a Thermo Scientific ESCALAB 250Xi XPS spectrometer and a monochromated Al K$\alpha$ X-ray source with energy 1486.6 eV. The measurement spot size was set to 500 $\mu$m and an electron flood gun was used to provide charge compensation. XPS analysis was performed using the Avantage software. All peaks were charged-corrected to the C 1s peak at 284.8 eV. The background signal arising from inelastically scattered electrons was removed using a Shirley baseline model and the peaks were fit using a Gaussian-Lorentzian product. 

{\textbf{{$\mu$SR} analysis addendum}}
\newline
TRIM.SP program was used to calculate muon stopping profile. It takes as an input parameter density of material and based on Monte Carlo simulations. For Nb$_2$O$_5$ we took 5.2 $\rm{g/cm^3}$, it is precise enough for our goals. For energy range up to 15 keV an average penetration depth of muon beam is sitting within 100 nm. 

\begin{figure}[!htb]
\includegraphics[width=0.7\textwidth]{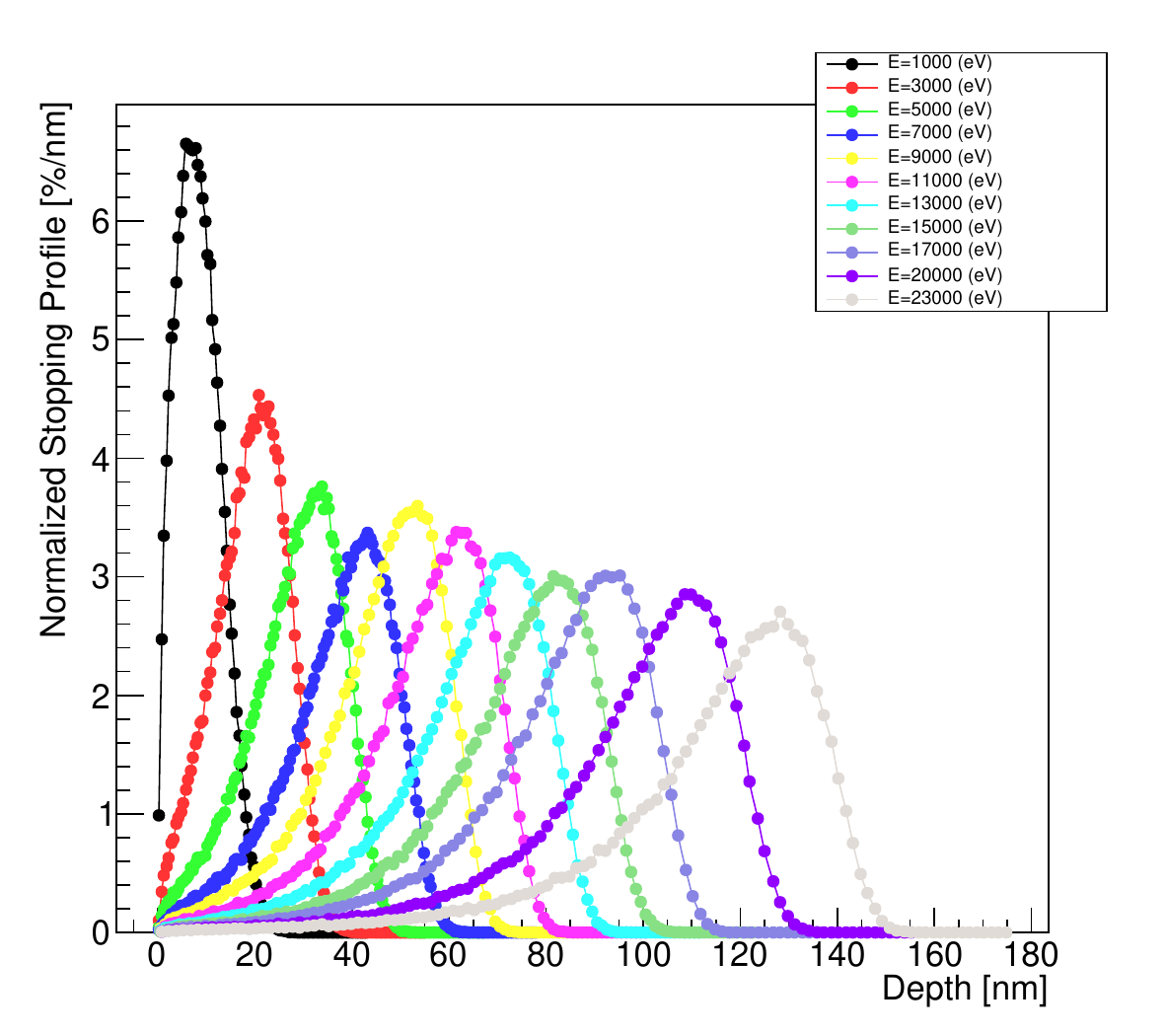}
\caption{\label{fig:profile} Simulated muon stopping profile for Nb$_2$O$_5$: muons' penetration depth.}
\end{figure}

\begin{figure}[!htb]
\includegraphics[width=0.5\textwidth]{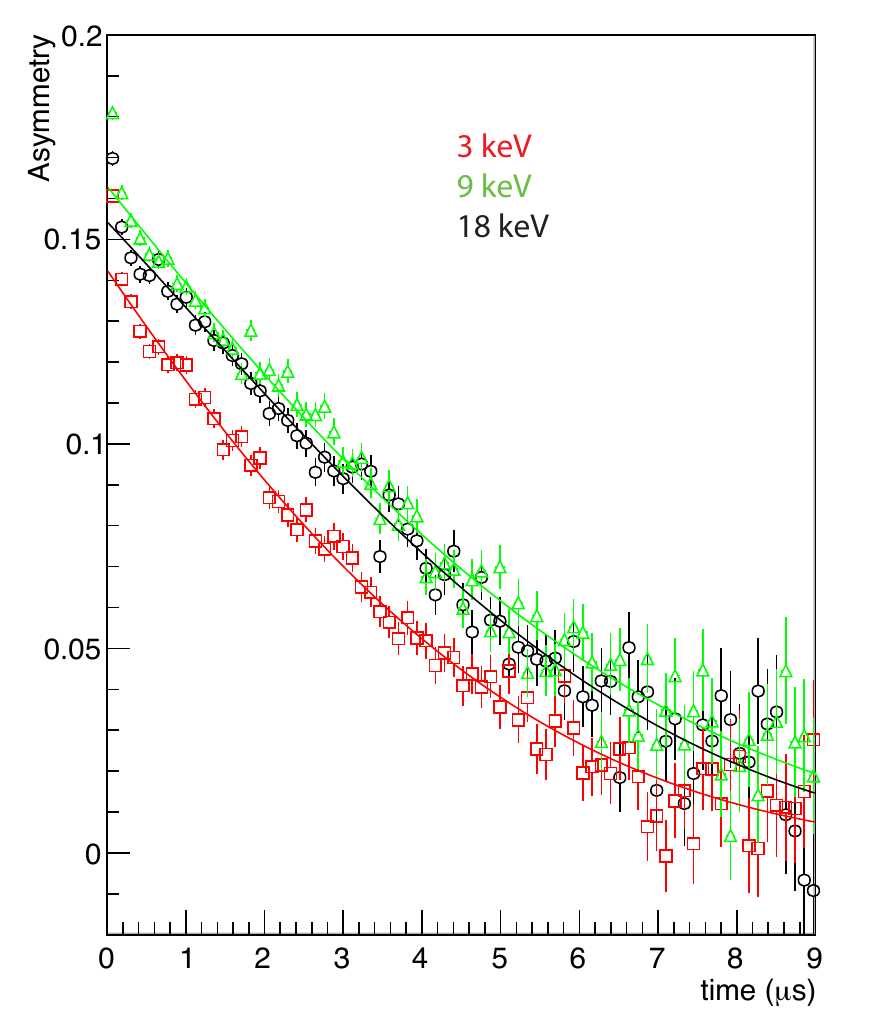}
\caption{\label{fig:ZF} $\mu$SR data in zero magnetic field (ZF) in Nb$_2$O$_5$, for different energies (penetration depths) $E= 3, 9, 18$~keV and base temperature $T=2.8$~K.}
\end{figure}

Zero field (ZF) muSR doesn't show significant difference for various penetration depth. It confirms that Nb$_2$O$_5$ film is uniform, and even more important is uniform from magnetism point of view.

\begin{figure}[!htb]
\includegraphics[width=0.7\textwidth]{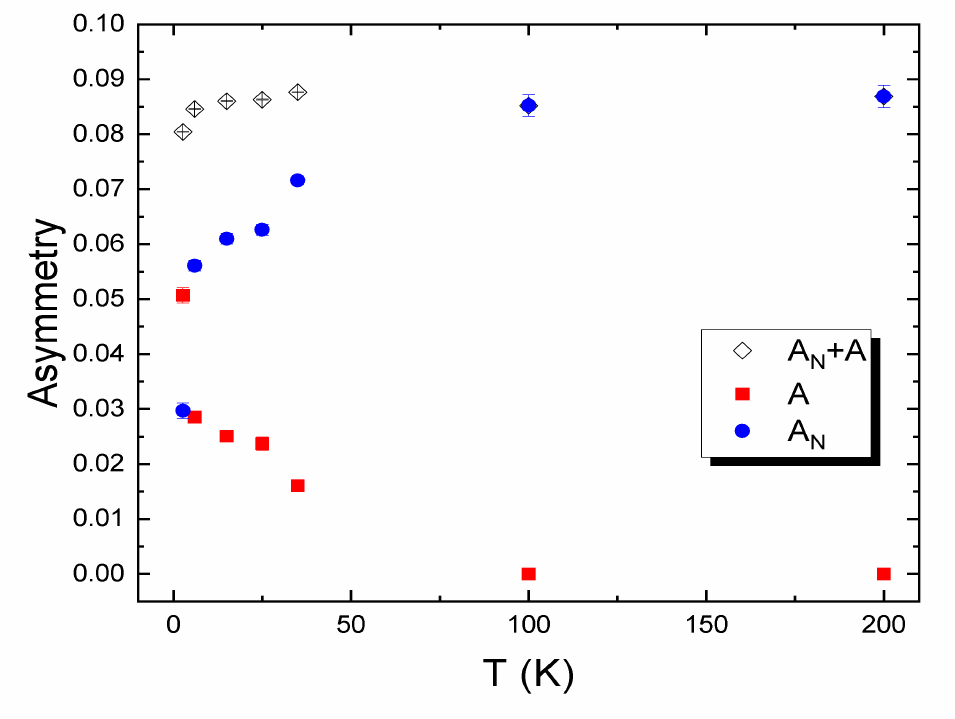}
\caption{\label{fig:asym_ne} Contributions of nuclear spins ($A_{N}$, blue points) and electron spins ($A$ red points) into asymmetry for Nb$_2$O$_5$, sum of inputs shown as black points. All data was obtained from global fit for variety of the longitudinal magnetic field (LF).}
\end{figure}

Asymmetry for LF measurements was extracted based on experimental data in the assumption that below 50~K temperature electronic system as a contributor to depolarization couldn't be neglected due to high fluctuation rate. Initial fitting took only one system approach and still demonstrated parameter $\nu$ significantly non zero in order of tens of MHz. Since applied field was high enough and nucliar system expected to be frozen. Notherless both nuclear and electronic inputs are taken into account during fitting of LF data using dynamical Kubo-Toyabe model.  Their asymmetry ratio is representing participation into muons' depolarization. One significant disadvantage that  model doesn't take into account interaction between electrons or potential clustering in mean field approach, so fitting doesn't give directly information about number of magnetic centers.

\nocite{*}

\bibliography{musrnboxcut}